\documentclass[aps,twocolumn,epsf,eps,floats,prl,showpacs, nofootinbib]{revtex4-1}
\usepackage{graphicx} 

\newcommand \be {\begin{equation}}
\newcommand \ee {\end{equation}}
\newcommand \bea {\begin{eqnarray}}
\newcommand \eea {\end{eqnarray}}
\newcommand \vv {\mathbf{v}}

\newcommand \rr {\mathbf{r}}

\newcommand \bs {\mathbf{s}}
\newcommand \hh {\mathbf{h}}
\newcommand \ppi{\mbox{\boldmath$\pi$}}
\newcommand \xxi {\mbox{\boldmath{$\xi$}}}
\begin{document}

\title{Boundary information inflow enhances correlation in flocking} 

\author{Andrea Cavagna$^*$, Irene Giardina$^*$, Francesco Ginelli$^{*\dagger}$}
\affiliation{$^*$ Istituto dei Sistemi Complessi, CNR, via dei Taurini 19, I-00185 Roma, Italy}
\affiliation{$^\dagger$ SUPA, Institute for Complex Systems and Mathematical Biology, King's College, University of Aberdeen, Aberdeen AB24 3UE, United Kingdom}

\date{\today} 

\begin{abstract}
The most conspicuous trait of collective animal behaviour is the emergence of highly ordered structures. Less obvious to the eye, but 
perhaps more profound a signature of self-organization, is the presence of long-range spatial correlations. Experimental data on 
starling flocks in $3d$ show that the exponent ruling the decay of the velocity correlation function, $C(r)\sim1/r^\gamma$, is extremely small,
$\gamma \ll 1$. This result can neither be explained by equilibrium field theory, nor by off-equilibrium theories and simulations of active 
systems. Here, by means of numerical simulations and theoretical calculations, we show that a dynamical field applied to the boundary 
of a set of Heisemberg spins on a $3d$ lattice, gives rise to a vanishing exponent $\gamma$, as in starling flocks. The effect of the
dynamical field is to create an information inflow from border to bulk that triggers long range spin wave modes, thus giving rise to an anomalously long-ranged correlation.
The biological origin of this phenomenon can be either exogenous - information produced by environmental perturbations is
transferred from boundary to bulk of the flock - or endogenous - the flock keeps itself in a constant state of dynamical excitation
that is beneficial to correlation and collective response.
\end{abstract}

\maketitle

%%%%%%%%%%%
%% INTRO
%%%%%%%%%%%

Flocking, the collective motion displayed by large groups of birds, is one of the most spectacular examples
of emergent collective behavior in nature, and it has fascinated inquiring minds since a long time \cite{Plinius}. 
 Statistical physicists have tackled the problem via 
minimal models of self propelled particles (SPP) \cite{Vicsek95, Gregoire2004} and hydrodynamic continuum 
theories \cite{TT, Ramaswamy, Bertin}. Such studies showed that flocking can be interpreted as a spontaneous 
symmetry breaking phenomenon occurring in a ``moving ferromagnetic spin system'', a sort of non-equilibrium 
counterpart of the well known Heisenberg model \cite{HRef}. The basic ingredients of this description -
 self propulsion, lack of Galileian invariance and momentum conservation, local ferromagnetic interactions - 
define an extremely rich universality class, able to describe systems as diverse as vertebrate herds \cite{Couzin2003}, 
bacteria colonies \cite{Swinney2010}, driven granular matter \cite{Chate2010}, grasshopper swarms 
\cite{Couzin} and active macromolecules in living cells \cite{Schaller2010, Sumino2012}.

Flocking, however, remains a prominent instance of collective animal motion for two reasons. First, it involves large
numbers of individuals, hence justifying a statistico-mechanical approach to the problem. Second, unlike for most 
$3d$ systems, for flocks of starlings ({\it Sturnus vulgaris}) we have experimental data \cite{Ballerini2008, Cavagna2008}, 
against which theories and models can be tested.  The statistical analysis of individual positions and velocities has revealed 
several unexpected physical features that need to be explained. In particular, it was found in \cite{Cavagna2010} that the 
spatial correlations of the velocity fluctuations in starling flocks are anomalously long-ranged. Such correlations are hard, if 
not impossible, to reconcile with the current theories of flocking.

Consider a flock of $N$ birds with velocities ${\bf v}_i$ and velocity fluctuations $\delta {\bf v}_i = {\bf v}_i - \frac{1}{N} 
\sum_i {\bf v}_i$. The two point correlation function is defined as,
\be
C(r,L) =\frac{\sum_{ij} C_{ij}\; \delta(r-r_{ij})}{\sum_{ij}
  \delta(r-r_{ij})} \ ,
\label{corr}
\ee
where $C_{ij}=\delta \vv_i\cdot\delta \vv_j$ and $r_{ij}$ is the distance between birds $i$ and $j$. In systems of finite size $L$, 
due to the global constraint $\sum_{i} \delta \vv_{i}=0$, the function $C(r)$ has a zero, which can be used as a finite-size definition 
of the correlation length $\xi$, $C(r=\xi)=0$. In starling flocks it was found that $\xi\sim L$, namely the correlation function is scale-free 
\cite{Cavagna2010}. In fact, long-range correlations are expected in systems where a continuous symmetry is spontaneously 
broken (the direction of motion for flocks, the spin direction for Hesemberg-like ferromagnets). 
However, in starling flocks correlations are {\it very} long-ranged. 

\begin{figure}[t]
\centering
\includegraphics[draft=false,clip=true,width=0.235\textwidth]{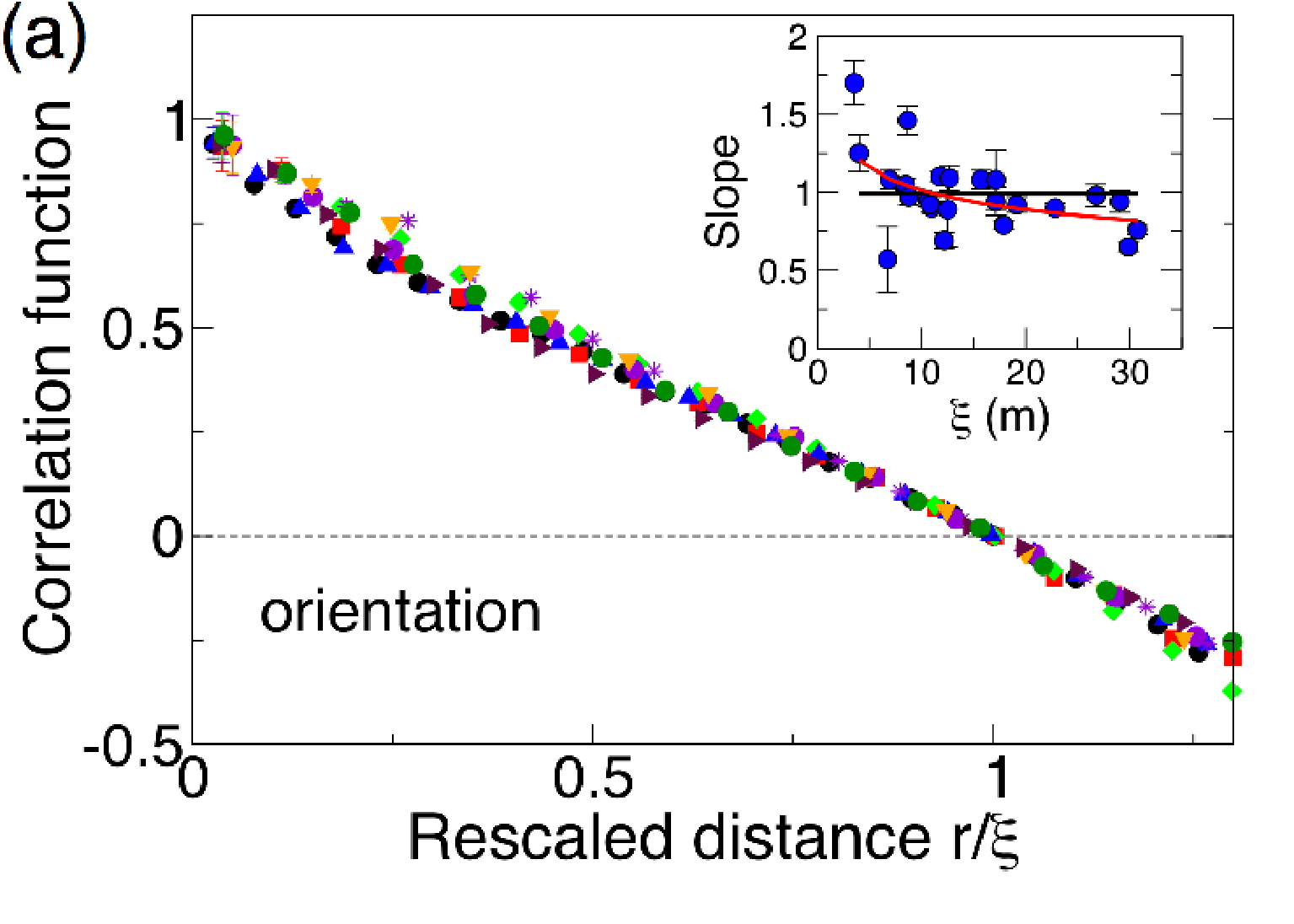}
\hspace{0.1 cm}
\includegraphics[draft=false,clip=true,width=0.23\textwidth]{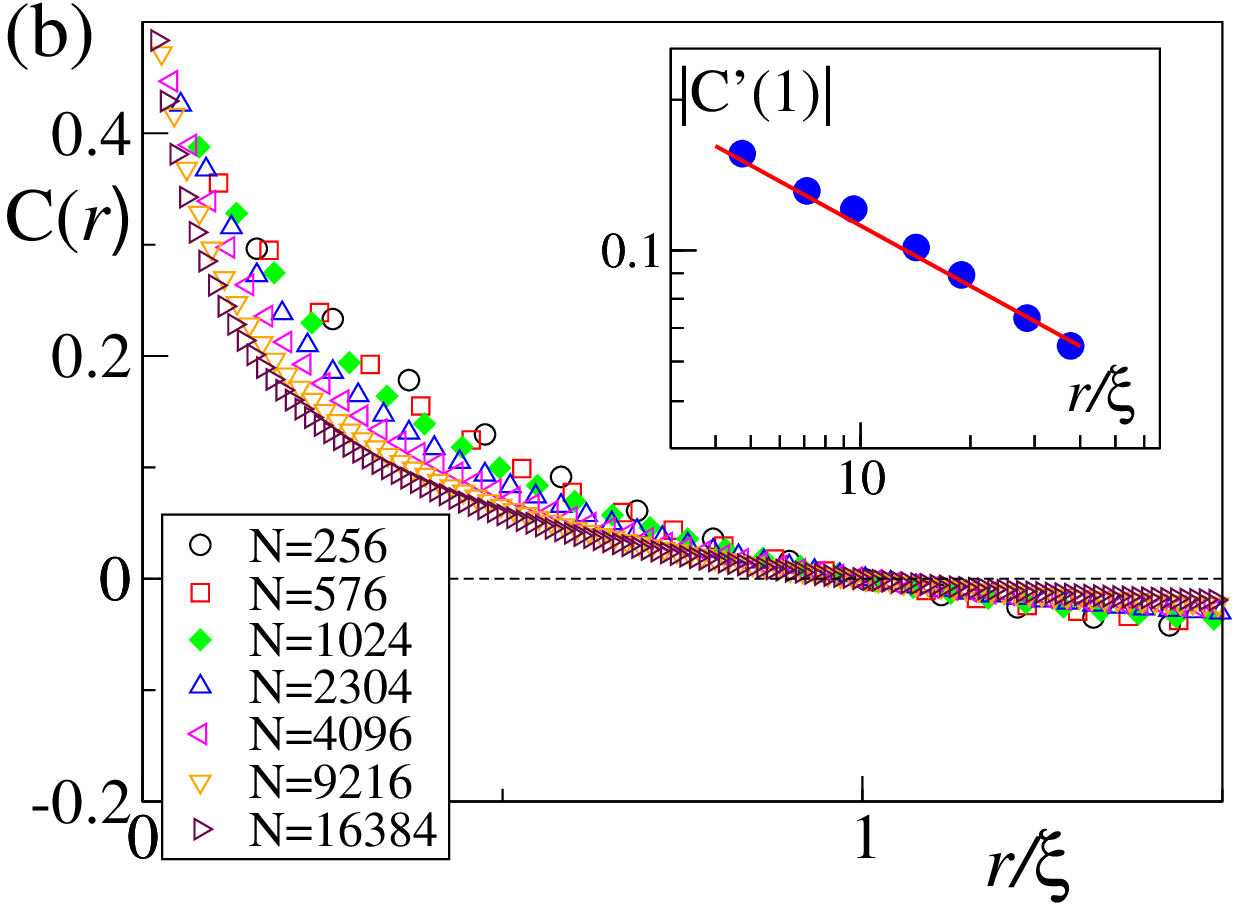}
\caption{(color online)
(a) Rescaled correlation functions in $3d$ experiments (a) and $2d$ flocking models (b).
Inset: modulus of the derivative
of the rescaled correlation function in $r/\xi=1$ vs. the correlation length
$\xi$. In both experiment and model, correlations are scale free, $\xi \sim
L$ (not shown here).
(a) 
 Experimental data from different highly ordered flocks with
different sizes (from 9.1m to 85.7m) and numbers of birds
(from 122 to 4268). (Reprinted from \cite{Cavagna2010})
(b)
Topological Vicsek model \cite{Chate2010}; numerical simulations in
the highly ordered regime on a $2d$ torus ($v_0=0.5$,
$N=256,\dots,16384$, angular noise with
amplitude $\eta=0.15$). Data averaged over $5\cdot 10^6$ timesteps. The inset fit (red line)
has slope $\gamma=0.4$.
}
\label{fig0}
\end{figure}

We can formalize the statement above in the following way. Let us write the 
finite-size correlation function as, $C(r,L)=\xi(L)^{-\gamma} g(r/\xi(L))$, where $g$ is a dimensionless scaling function, with $g(1)=0$ \cite{scaling}. 
Hence, the derivative in $1$ of the rescaled correlation function, $C(r/\xi)$, is given by $C'(r/\xi=1)\sim \xi(L)^{-\gamma}\sim L^{-\gamma}$.
In \cite{Cavagna2010} it was found that $C'(1)$ does not show any 
significant scaling with $L$ (or, equivalently, with $\xi$) (Fig.~\ref{fig0}a), implying $\gamma \sim 0$ \cite{NOTE}. 
This fact is surprising, as it implies that in two flocks of sizes $L$ and $2 L$, one gets $C(2 r, 2 L) \sim C(r, L)$. 
Correlations are basically not decaying at all.

This result  contrasts with the classical Heisenberg model on a $3d$ lattice, where $\gamma=1$, implying
$C(2 r, 2 L) \sim 1/2 \; C(r, L) $ \cite{HRef}. The situation does 
not improve when one considers non-equilibrium flocking theories and
models. While different scalings are expected in the perpendicular and
parallel directions with respect to the mean velocity, the
hydrodynamic approach of \cite{TT} predicts to leading order $\gamma=6/5$ in $3d$
and $\gamma=2/5$ in $2d$ \cite{Toner2012}, a result supported by numerical simulations
for the $2d$ topological SPP model introduced in \cite{Ginelli2010} (Fig.~\ref{fig0}b). 

Therefore, both in $3d$ and $2d$, SPP models and hydrodynamic theories predict a decay of 
the correlation function that is {\it faster} than the equilibrium case, whereas in starling flocks one finds a 
decay dramatically {\it slower} than the equilibrium case. We conclude that the origin of the anomalously 
slow decay of correlation in starling flocks is probably not to be found in the self-propelled nature of real birds.
The discrepancy between models and theories on one side, and bird flocks on the other side, is troublesome. It has been 
suggested in \cite{Cavagna2010} that these unusually strong correlations may be responsible for the very effective  
response of flocks to external perturbations, most notably predators attacks. If this is true, it means that the value of $\gamma$ may play
a relevant evolutionary role. Hence, understanding what is going on seems important.

A first hint about the origin of this phenomenon was given in \cite{Bialek}. There, it was shown that the minimal model inferred from the data 
via a maximum entropy criterium, correctly reproduces the slow decay of the correlation function only if velocities on the flock's boundary are 
kept fixed to their experimental value, while the bulk velocities follow the model dynamics. 
This result suggests that the slow decay of the correlation function
could be caused by an information transfer from the boundary to the
bulk of the flock.  However, it is not {\it a priori} clear which
mechanisms is able to enhance correlations in such a dramatic
way. Here, we hypothesize that this could be due to a dynamic
information inflow, and test this conjecture by studying the dynamics of a finite-size Heisenberg ferromagnet, 
under the effect of a dynamical magnetic field that affects a part of the boundary. 

\begin{figure}[t]
\centering
\includegraphics[draft=false,clip=true,width=0.47
\textwidth]{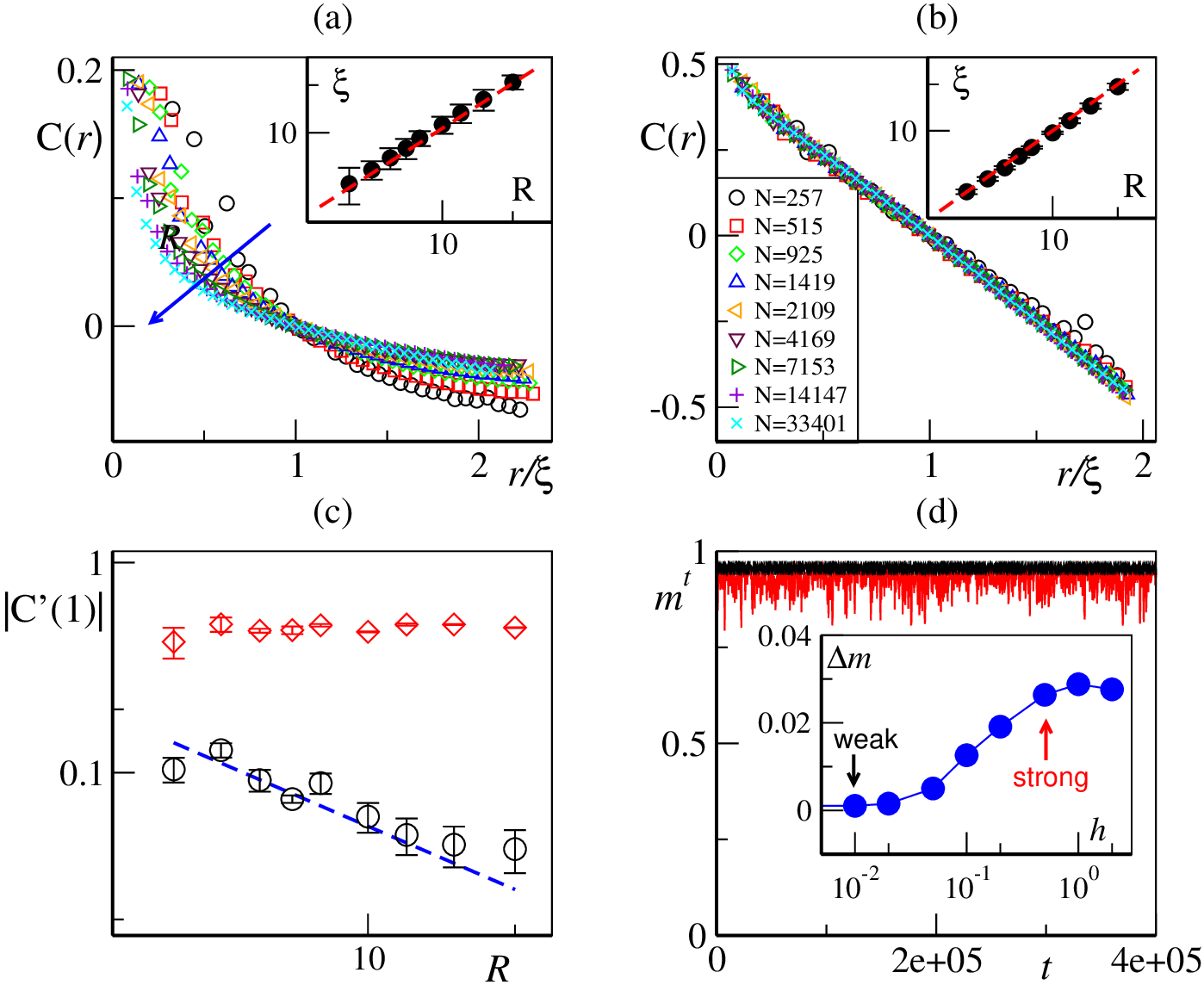}
\caption{(color online) Time averaged, two points correlation of spin
fluctuations ($\eta=0.3$, $\alpha=2$). 
(a)-(b) Correlation function $C$ vs. the rescaled distance
$r/\xi$ for system radii $R$ from 4 to 20 (total number of spins in
the legend): (a) weak field $h=0.01$ and (b) strong field $h=0.5$. 
In the insets: correlation length $\xi$ vs. $R$ in log-log scale. 
The dashed red lines mark
linear growth.
(c) Log-log plot of the rescaled correlation function slope at
$r/\xi=1$ vs. system size $R$ for weak (black circles) and strong (red
diamonds) fields. The dashed blue line marks the decay as $1/R$.
%extpected in the zero field case.
(d) Magnetization $m^t$ timeseries for the above weak (black line) and strong
(red line) field cases for $R=15$. Inset: Its standard deviation $\Delta m$
as a function of field strength. 
}
\label{fig2}
\end{figure}
%

%%%%%%%%%%%%%%
%% numerical simulations
%%%%%%%%%%%%%%

We consider the classical Heisenberg model with nearest neighbour interactions, defined on a 
spherical portion (of radius $R$) of a $3d$ cubic lattice.
The boundary $\mathcal{B}$ of the sphere (defined as the set of spins with less than 6 neighbors)
is affected by a dynamical external field $\hh^t$ which keeps the system out of equilibrium and 
determines an information flow from the boundary to the bulk. The external field has fixed modulus $h$, is outward pointing 
and at each time step is applied to only half of the spherical boundary.  The field dynamics follows a 
uniform random walk on a sphere, designed in such a way that $\hh^t$ reverses its direction on average in a time $\tau_h=R^{\alpha}$ 
\cite{NOTE2}. 

Spins $\bs_i^t$ are unitary vectors with lattice coordinate $\rr_i=(x_i,y_i,z_i)$, where $i=1, \ldots, N$ and $x_i,y_i,z_i$ are integers such that
$r_i=||\rr_i|| \leq R$. Spins follow the time discrete, synchronous dynamics,
\be
\bs_{i}^{t+1}=\Theta \left[\Theta \left[\bs_{i}^t + \sum_{j \in i} \bs_{j}^t + {\bf g}(\rr_i, \hh^t) \right] + \eta {\bf \zeta}_{i}^t \right]  \ ,
\ee
where the sum runs over the lattice nearest neighbors of $\rr_i$, $\theta[\vv]=\vv/||\vv||$ is a normalization operator 
and ${\bf \zeta}_{i}^t$ is a random vector delta correlated in space and time and uniformely distributed 
in the unit spherical surface. Boundary conditions are determined by the function
${\bf g}(\rr_i, \hh^t)=\hh^t$ if $\rr_i \in \mathcal{B}$ and $(\hh^t \cdot \rr_i) > 0$,
while ${\bf g}(\rr_i, \hh^t)=0$ otherwise.
For zero field, $\hh^t=0$, the dynamics converges towards the equilibrium distribution of an Heisenberg ferromagnet
with a temperature $T$ that is a monotonic function of the noise amplitude $\eta$ \cite{discrete_dynamics}.

Being interested in the highly ordered phase, we fix
noise to $\eta=0.3$. We initially consider $\alpha=2$, so that the typical field inversion time
$\tau_h = R^2$ is of the order of the information propagation time as given by standard diffusive dynamics.
In order to compare with the results of \cite{Cavagna2010}, we define spin fluctuations  as ${\bf u}_{i}^t= \bs_i^t-{\bf m}^t$, 
where ${\bf m}^t\equiv m^t {\bf n^t}=\frac{1}{N} \sum_{i} \bs_{i}^t$ is the instantaneous
global magnetization and ${\bf n^t}$ its unitary direction. The correlation function is defined as in 
(\ref{corr}), with $C_{ij}=\langle {\bf u}_{i}^t \cdot {\bf u}_{j}^t \rangle_t$, where $\langle \cdot\rangle_t$
denotes a time average over a scale $\tau >> \tau_h$ \cite{NOTE3}.
  
The effect of a strong dynamical field on the correlation function is striking. In Fig.~\ref{fig2}a-b, we report $C(r)$
for different `flock' sizes $R$, at two values of the field, $h=0.01$ and $h=0.5$. In both cases the correlation length $\xi$ 
grows linearly with $R$, as expected in a scale-free system as Heisenberg. Moreover,  in the weak field case the 
rescaled correlation function $C(r/\xi)$ behaves as in the equilibrium case: the derivative of the correlation function 
at $r/\xi=1$, vanishes for increasing sizes, $C'( 1)\sim 1/R^\gamma$, with $\gamma=1$ (Fig.~\ref{fig2}c, black circles). 
On the contrary, in the strong field regime the correlation has a striking resemblance with that observed in real bird flocks 
\cite{Cavagna2010}. In particular, the correlations $C(r/\xi)$ at different sizes collapse onto a single curve. 
This means that in the strong field regime, the derivative $C'(1)$ is constant, implying $\gamma \sim 0$ (Fig.~\ref{fig2}c, red diamonds), 
in agreement with experiments (Fig.\ref{fig0}a).

Having enhanced correlation, we must make sure we have not destroyed order. Hence, let us 
check the field effects on the global magnetization, which is the equivalent of flock's velocity. 
In Fig.\ref{fig2}d we show the time series of the scalar magnetization $m^t$ for weak (black) and strong field (red). 
While the strong field standard deviation $\Delta m$ is increased by about a factor 20, the mean magnetization 
is only slightly reduced, so that the ferromagnet remains in the deeply ordered phase. Note also that the variance 
saturates as the field is increased past $h=0.5$, so that magnetization fluctuations stay
finite and relatively small even in the strong field regime.

\begin{figure}[t]
\centering
\includegraphics[draft=false,clip=true,width=0.2\textwidth]{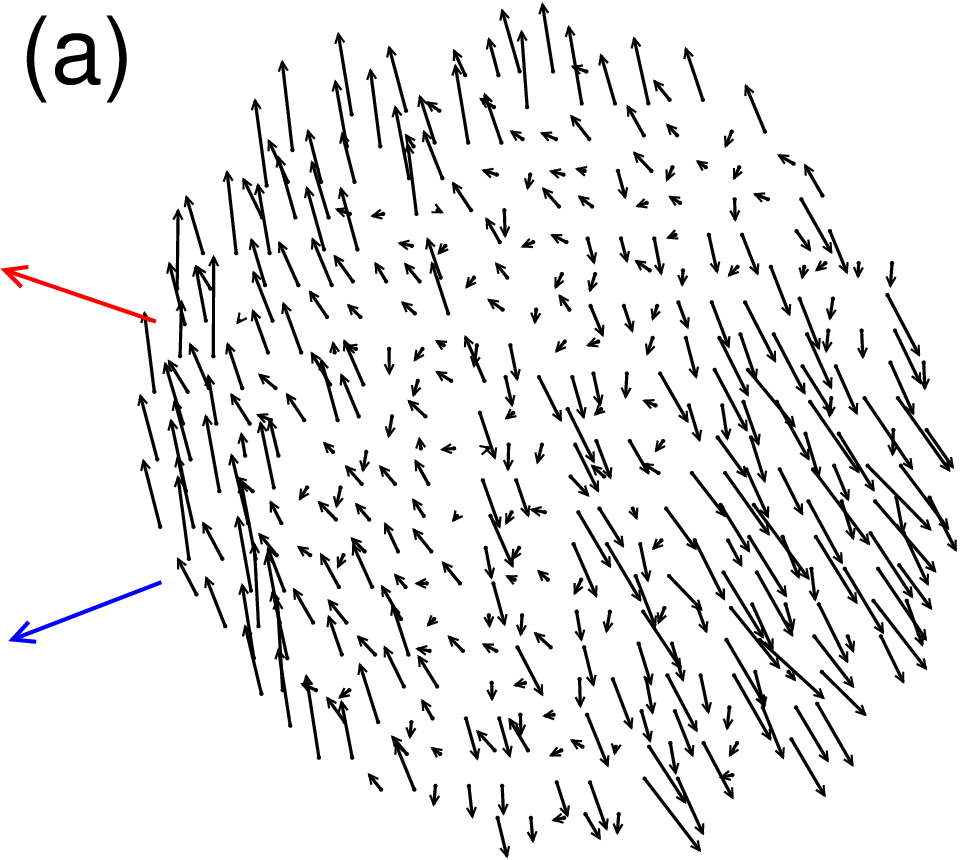}
\hspace{0.2 cm}
\includegraphics[draft=false,clip=true,width=0.2\textwidth]{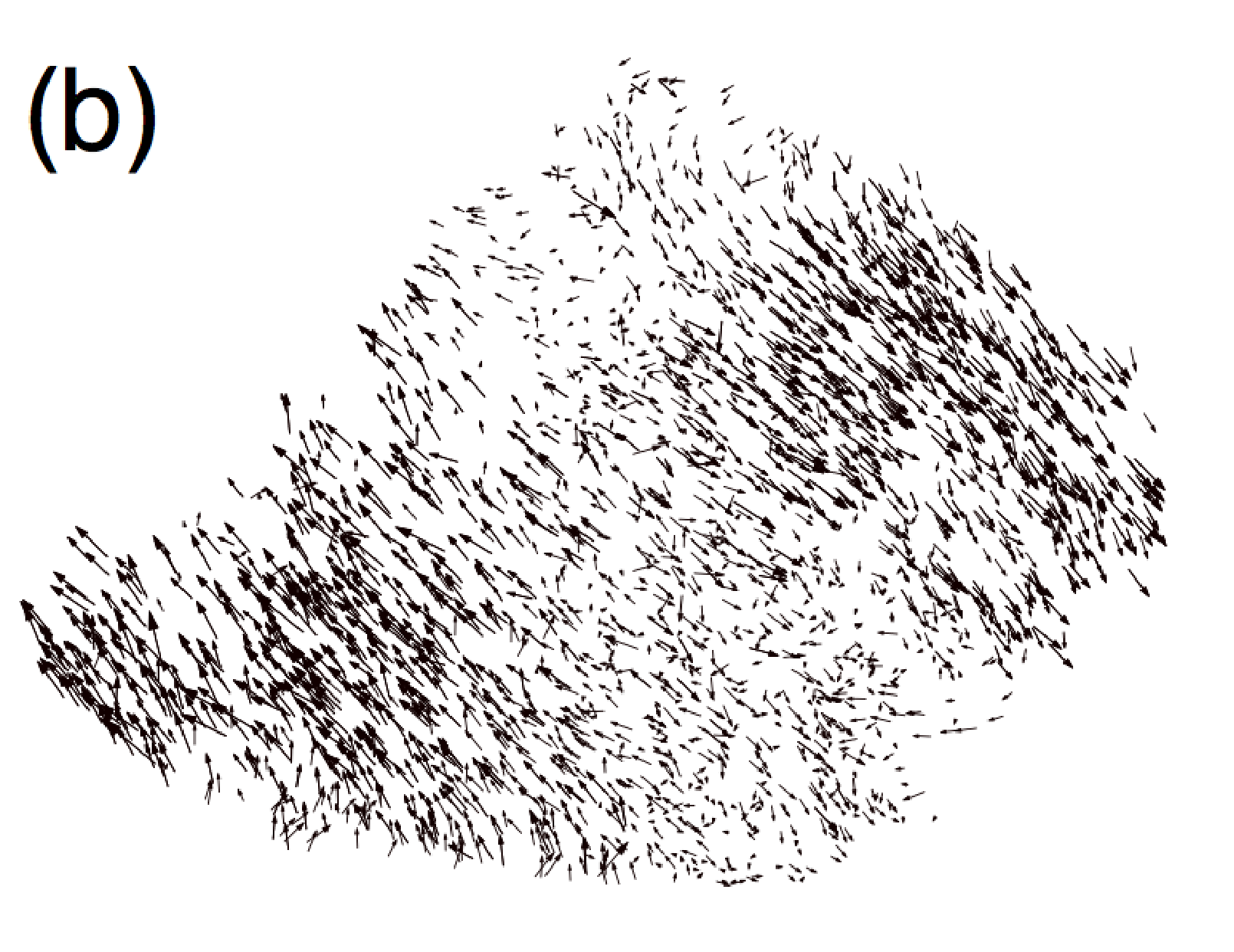}
\caption{(color online) (a) Typical snapshot of spin fluctuations in the field (red arrow)-magnetization (blue arrow) plane
for $R=8$, $h=0.5$, $\eta=0.3$, $\alpha=2$.
The arrows length has been rescaled for clarity reasons.
(b) Typical snapshot of velocity fluctuation in a starling flock. (Reprinted from \cite{Cavagna2010})
}
\label{fig3}
\end{figure}

%%%%%%%%%%%%%%%%%%%%%%%%%
%{\it Theory in spin wave approximation}\\
%%%%%%%%%%%%%%%%%%%%%%%%%

We can analytically explain our numerical result by using the spin-wave approximation. Let us start by considering the Heisenberg model 
at equilibrium. Each spin can be decomposed as  ${\bf s}_i=s_i^L {\bf n} + \ppi_i$, where
$s_i^L$ and $\ppi_i$ are the longitudinal and perpendicular components with respect to magnetization. 
At low temperature, when the system is highly polarized, one has
$\pi_i^2\ll 1$, so that, using the unitary condition $||{\bf s}_i||=1$, 
we get, $s_i^L\sim 1-\pi_i^2/2$ (we also note that  ${\bf u}_i=\ppi_i$ at leading order). Under these conditions 
the original Hamiltonian, ${\cal H} = - 1/2 \sum_{\langle i,j\rangle} {\bf s}_i\cdot {\bf s}_j$, can be expanded, leading to a Gaussian partition function,
\begin{equation}
Z\sim \int D{\ppi} \; \delta\left(\sum_i{\ppi}_i\right) \exp{\left\{-\frac{\beta}{2} \sum_{ij} A_{ij} {\ppi}_i\cdot{\ppi}_j\right\}}   \  ,
\label{Z-sw}
\end{equation}
where, $A_{ij}=\sum_k n_{ik} -n_{ij}$, is the discrete Laplacian, and the adjacency matrix $n_{ij}$ is $1$ for nearest neighbors, and $0$ otherwise.  
The connected correlation function $C_{ij}$ can be easily computed in terms of the eigenvalues $\{\lambda_a\}$ and eigenvectors $\{{\bf w}^a\}$ 
($a=1\cdots N$) of $A_{ij}$ \cite{Bialek}, 
\begin{equation}
C_{ij} = \langle \ppi_i\cdot\ppi_j \rangle
= \sum_{a>1} w^a_i w^a_j \frac{2}{\beta \lambda_a}  \ ,
\label{correlation-sw}
\end{equation}
where $w^a_i=\langle {\bf w}^a | {\bf i}\rangle$, in Dirac's notation.
The matrix $A$ has a zero eigenvalue related to the original rotational symmetry of the 
Hamiltonian. The first non-zero eigenvalue is of order $1/R^2$ on a discrete lattice of size $R$;
it is indeed the presence of this soft (or massless) mode that gives rise 
to long-range correlations when $R\to\infty$ \cite{HRef}. 

Let us first study the effect of a {\it static} field, ${\cal
  H}\rightarrow {\cal H} -\sum_i {\bf s}_i \cdot {\bf h}_i$. The field breaks the rotational symmetry and perturbs the diagonal part of the Laplacian matrix, $A_{ii}\rightarrow A_{ii}(h)=\sum_{k}A_{ik} + {\bf h}_i\cdot{\bf n}$. To first order we have, $\lambda_a \rightarrow \lambda_a(h) = \lambda_a(0)+\sum_i (w^a_i)^2 \, {\bf h_i}\cdot {\bf n}$. If the field is homogenous and acts over $O(R^3)$ sites,  this correction is $O(1)$: eigenvalues are shifted by a nonzero mass and correlations are no longer scale-free. On the other hand, 
if the field only acts on a number of sites of $O(R^2)$ (as the boundary), then we get, $\lambda_a(h)\sim \lambda_a(0)+O(1/R)$, and correlations remain long-range. Hence, the first thing we learn
is that the field must not be applied on all spins lest correlation becomes short range.

From (\ref{correlation-sw}) we also learn that the correlation function is a superposition of normal modes ${\bf w}^a$ (the so-called spin waves). 
Each one of these modes has a specific space modulation (on a cubic lattice they are plane waves). 
The lowest non-zero modes correspond to fluctuations with length-scale $R$ that reverse the orientation of the spins from one side of the system to the other, very similar to the
fluctuations observed in real flocks (Fig~\ref{fig3}). Hence, to lower the value of $\gamma$ we must apply a (boundary) field that overweights these 
long-range modes. Since a static field leads to $\gamma=1$ at best, it seems natural 
to consider a time-dependent boundary field, ${\bf h}(t)$.

To treat this case we consider the  Langevin equation,
\begin{equation}
\frac{d{\bf s}_i}{dt}=\sum_{j} n_{ij} {\bf s}_j + {\bf h}_i-\mu_i {\bf s}_i + \xxi_i  \ ,
\label{langevin}
\end{equation}
where $ \mu_i$ is a time-dependent Lagrange multiplier enforcing the constraint $||{\bf s}_i|| =1$, and  $\mbox{\boldmath{$\xi$}}_i$ is a vectorial delta-correlated noise. As in our
numerical simulation, we choose the subset $\cal B$ where the field is applied to be a fraction of the boundary, ${\bf h}_i(t) = {\bf h}$ for $i\in {\cal B}(t)$, with $\bf h$ directed outward.
The field has a timescale $\tau_h\sim R^\alpha$. If temperature is low and the field varies slowly in time, we can describe the system in terms of the polarization direction ${\bf n}(t)$ and of 
the instantaneous perpendicular fluctuations $\{\ppi_i(t)\}$ around it.  Projecting Eq.(\ref{langevin}) along and perpendicularly to ${\bf n}$, and exploting the equation for the constraint $\mu_i$, we get,
\begin{eqnarray}
\frac{d\ppi_i}{dt}&=&\!-\!\sum_{j}A_{ij} \ppi_j\! -\!({\bf h}_i\cdot \ppi_i)\ppi_i\! +\! {\bf h}_i^\perp \!+\! \xxi_i^\perp \phantom{pppp}\nonumber\\
&&\phantom{pppppppppp} - (1-\frac{\ppi_i^2}{2})\frac{d{\bf n}}{dt}-(\ppi_i\cdot \frac{d{\bf n}}{dt}){\bf n}  \ ,  
\label{dyn-pi}
\\
M \frac{d {\bf n}}{dt}&=&-\frac{({\bf h}\cdot{\bf n})}{N}\sum_{i\in {\cal B}} \ppi_i -\frac{1}{N}\sum_{i\in {\cal B}}({\bf h}\cdot \ppi_i)\ppi_i +\frac{B}{N} {\bf h}^\perp\phantom{pppp}  
\label{dyn-n}
\end{eqnarray}
where $B$ is the cardinality of ${\cal B}$, ${\bf h}_i=h_i^L {\bf n} +
{\bf h}_i^\perp$ and the perpendicular
component of the noise $\xxi_i^\perp$ has variance $4T$.
\begin{figure}[t]
\centering
\includegraphics[draft=false,clip=true,width=0.45\textwidth]{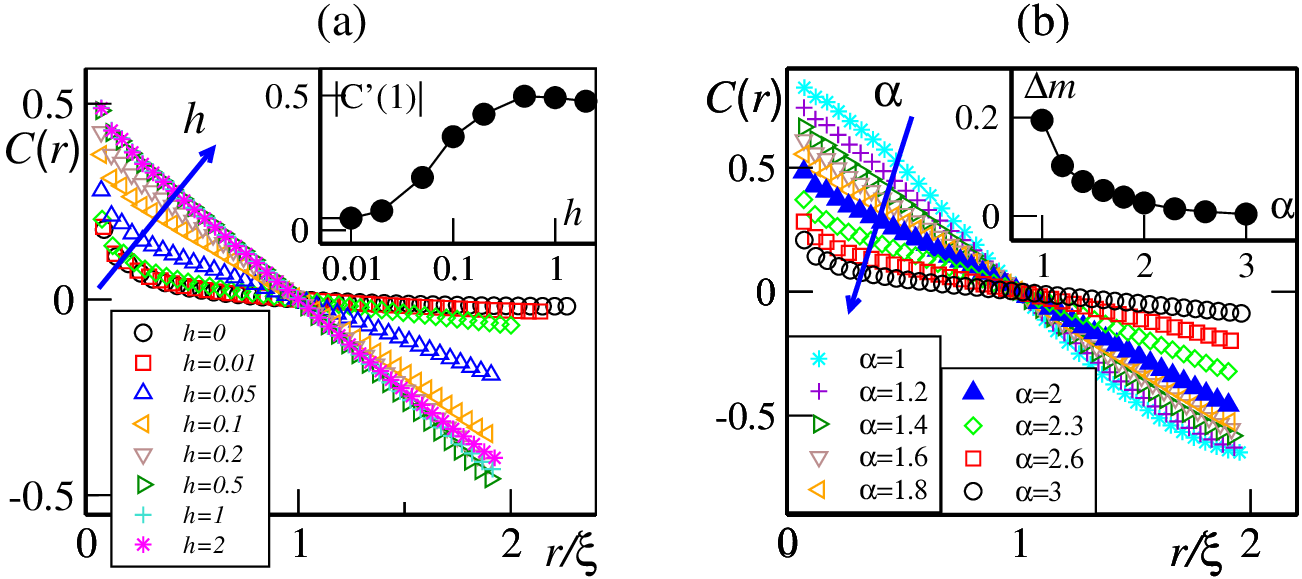}
\caption{(color online) (a) Correlation functions for different external field strength (increasing in the
blue arrow direction) at fixed system size 
$R=15$ and $\alpha=2$. In the inset, rescaled correlation function slope at $r/\xi=1$ vs. $h$ in a semi-log scale.
(b) Correlation functions for different field dynamical exponent $\alpha$ (increasing in the
blue arrow direction) at fixed system size $R=15$ and field strength $h=0.3$. Full symbols
refer to the ``diffusive'' exponent $\alpha=2$.
In the inset, magnetization standard deviation as
a function of $\alpha$. In both panels $\eta=0.3$.
}
\label{fig4}
\end{figure}
From these equations we can gain information on how the fluctuations behave. If we assume that $\bf n$ does not change significantly on the scale over which fluctuations decay, then the leading behaviour of the correlation function can be estimated using the first line of Eq.(\ref{dyn-pi}). One gets,
\begin{eqnarray} 
&&C_{ij}(t)= \sum_{a,b} w^a_i w^b_j \left \{ C^0_{ab} e^{-(\lambda_a+\lambda_b)t} + \delta_{ab} \frac{2T}{\lambda_a} 
\left ( 1-e^{-2\lambda_a t}\right ) \phantom{}\right.\nonumber \\
&& \phantom{}\left. + 2 \int_0^t dt' e^{-(\lambda_a+\lambda_b)(t-t')}\left [ {\bf m}_a(t')\cdot{\bf h}^\perp_b(t')
%+{\bf m}_b(t')\cdot{\bf h}^\perp_a(t')
\right ]\right\}+b_{ij}  \ ,
\label{corre-off}
\end{eqnarray}
where $b_{ij}$ is a renormalization term enforcing the constraint, the variables with index $a$ indicate a projection on the $a$ normal mode (e.g. ${\bf h}_a = \sum_i w^a_i {\bf h}_i$), and the magnetization ${\bf m}_a(t)$ is given by,
\begin{equation}
{\bf m}_a(t)={\bf m}_a^0e^{-\lambda_a t} + \int_0^t dt' e^{-\lambda_a(t-t')}{\bf h}^\perp_a(t') \ .
\end{equation}
The first term on the r.h.s. of Eq.(\ref{corre-off}) depends on the initial conditions; the second term is the dynamical counter-part of the standard Heisenberg correlation and would be present even in absence of any external field; the last term, on the contrary, is the one mainly affected by the presence of a field and by its dynamics. We can see that each mode ${\bf w}^a$ gives a contribution to the correlation decaying on a time $\tau_a\sim1/\lambda_a$. If the field time-scale, $\tau_h$, is much larger than the maximum $\tau_a$ (slowest mode), the field is as good as constant, and fluctuations equilibrate to their static expression. In this case, correlations are of the Heisenberg kind, $\gamma=1$.

However, if $\tau_h$ is in the same range as the spin wave time scales, all the modes with $\tau_a>\tau_h$ do not equilibrate. In particular, we recall that the slowest modes has $\tau_a\sim1/\lambda_a\sim R^2$, and therefore by choosing $\tau_h \sim R^2$, we make the faster modes relax, but we keep the lowest modes excited. The last contribution in Eq.(\ref{corre-off}) is therefore non-trivial and - if stronger than the second standard term - it can modify the large scale behaviour of the correlation. For this to occur, we need $h^2_a\gg \beta \lambda_a$. Given that ${\bf h}_a = \sum_{i\in {\cal B}} w^a_i {\bf h}_i$ and that $(w^a_i)^2 \sim 1/N$ (${\bf w}^a$  has norm $1$), we obtain that the `strong-field' regime is defined by the condition $h^2 B^2/N > \beta/R^2$, i.e. $T R^3 h^2 >  1$. 

This last result tells us that the definition of strong vs weak field depends on the size $R$.
For example, for $h= 0.01$ (black circles in Fig.\ref{fig2}c) there should be no relevant effect of the dynamical field
for $R<10$ (weak field), while some field-induced departure from $\gamma=1$ should be visible for $R>10$ (strong field),
which is indeed what we see in Fig.\ref{fig2}c. On the contrary, for $h=0.5$ (red diamonds in Fig\ref{fig2}c), $\gamma=1$ is 
violated at as low a size as $R \sim 1$. In Fig.\ref{fig4}a we explicitly show the effect of crossing over from weak to strong field.  

%%%%%%%%%%%%%%%%%%%%%%%%%
%% test numerico delle predizioni teoriche
%%%%%%%%%%%%%%%%%%%%%%%%%

We finally checked numerically the effect of changing the field timescale, $\tau_h$ (Fig.\ref{fig4}b). Slow fields ($\alpha>2$) produce
correlation functions very close to the equilibrium case, while for $\alpha \leq 2$ correlations approach the linear decay  observed in bird flocks. 
Therefore, a random dynamical field evolving on a scale faster than $R^2$ enhances correlations in our spin system.
However, the effect of such fast fields can be detrimental to global order. In the inset of Fig.\ref{fig4}b
it is shown that the standard deviation of the magnetization increases sharply as $\alpha \to 1$, so that fields evolving too fast effectively destroy order in the system, whereas dynamics near the timescale, $\tau_h \sim R^2$, preserves order {\it and} increases correlations. 

We note that the optimal timescale, $\tau_h \sim R^2$, 
exactly characterizes {\it diffusive} information propagation in the
Heisenberg lattice. However, in real flocks birds seem to move in the
center of mass reference frame in a way closer to ballistic than to
diffusive, $\delta r^2 \sim t^{1.7}$ \cite{silvio, EPJB}.
It is therefore possible that the `right' timescale for enhancing the
correlation in natural flocks should be somewhat faster
(i.e. $\alpha<2$) than the purely diffusive one.
  
Velocity fluctuation correlations have been recently measured also in $2d$ colonies of motile bacteria ({\it Bacillus subtilis}) \cite{Bacteria}. This study 
finds scale-free correlations, as in the case of bird flocks. However, at variance with flocks, the decay exponent found for bacteria ($\gamma \in [0.1,0.2]$), is not
in plain contradiction with existent theories. As we said, the hydrodynamic approach in $2d$ gives $\gamma=2/5$ \cite{TT}, not far from $0.2$, considering experimental error.
Moreover, the linearized hydrodynamics theory, which is expected to describe small clusters, and thus to be more suited to the data of \cite{Bacteria}, 
predicts $\gamma=0$ in $2d$ \cite{TT}.
% again not too far from the experimental value. 
It therefore seems that in the case of bacteria, data can
be explained without the need of the border perturbation theory developed here.

%%%%%%%%%%%%%%%%%%%%%%
% CONCLUSIONS
%%%%%%%%%%%%%%%%%%%%%%

% main result

We have shown that a dynamical information flow due to a
fluctuating field and propagating from boundary to bulk, 
gives rise to strong correlations akin to the ones observed 
in real flocks. What is the biological origin (if any) of such information inflow? 
The border of a flock is hit by an ever changing flux of environmental stimuli and perturbations: attacking falcons, disturbing seagulls, wind gusts, sight of 
significative landmarks, are just few examples. These stimuli could give rise to the strong observed correlation. 
A few caveats arise about this {\it exogenous} hypothesis. 
It is unclear whether external stimuli can yield a time scale at least comparable to the optimal one, $\tau_h(R)$, for every biologically available 
flock size $R$. To check this point it would be important to study the effect of a perturbing field  characterized 
by several different time scales, and see whether the optimal
time-scale that enhances correlation is naturally selected by the
system. Another issue is that there are times at which flocks seem not
to be subject to evident dynamical perturbations, and yet display the same anomalous correlations. 

An alternative hypothesis is that the origin of the phenomenon is {\it endogenous}: even in absence of environmental perturbations, the flock sets itself 
constantly into a state of dynamical excitation because this behaviour 
%has the effect to 
enhances correlation and collective response when true perturbation 
strikes. Within this scenario, correlation is the evolutionary {\it cause} of the dynamical excitation, not the {\it by-product} of it. Moreover, the time 
scale of the endogenous excitation would be naturally related to the flock's size $R$, as it is the very birds on the border that spontaneously 
create the excitation. The problem with this endogenous hypothesis is how the flock would do that. New models, able to account for 
spontaneous states of dynamical excitation, may help understanding this point.

This work was supported by grants IIT Seed Artswarm, ERCÐStG n.257126, and AFOSR Z80910.

%%%%%%%%%%%%%%%%%%%%%%%%%%


\begin{thebibliography}{99}
\bibitem{discrete_dynamics}
The time-discrete dynamics preserves the underlying equilibrium measure and has a twofold advantage:
it simplifies numerical simulations;  it is the lattice counterpart of SPP models, 
which describe a driven overdamped dynamics and are thus intrinsically time-discrete.
\bibitem{NOTE2}
In spherical coordinates $(\theta, \phi)$, we draw the angular steps
$\Delta \phi = \chi_1\frac{\pi}{4}\frac{\Delta h}{\cos \theta} $ and
$\Delta \theta = \arcsin (\sin \theta -  \chi_2 \Delta h/ 2) - \theta$, where
$\Delta h = R^{-\alpha/2}$ is the random walk stepsize and $\chi_1$, $\chi_2$ are
two random numbers, delta correlated in time and uniformely distributed in $[-1,1]$.
\end{thebibliography}
\end{document}